\documentclass[a4paper,11pt]{article}
\pdfoutput=1 

\usepackage{jcappub,color} 

\usepackage[T1]{fontenc} 
\newcommand{\black}{\color{black}}
\newcommand{\red}{\black}

\title{Supermassive dark-matter Q-balls in galactic centers?}

\author[a,b]{Sergey Troitsky}
\affiliation[a]{Institute for Nuclear
Research of the Russian Academy of Sciences,\\
60th October Anniversary
Prospect 7a, Moscow 117312, Russia}
\affiliation[b]{
\red
Moscow Institute for Physics and Technology,\\
Institutskii per.\ 9, 141700, Dolgoprudny, Moscow Region, Russia
}

\emailAdd{st@ms2.inr.ac.ru}

\abstract{
Though widely accepted, it is not proven that supermassive compact objects
(SMCOs) residing in galactic centers are black holes. In particular, the
Milky Way's SMCO can be a giant nontopological soliton, Q-ball, made of a
scalar field: this fits perfectly all observational data. Similar but tiny
Q-balls produced in the early Universe may constitute, partly or fully, the
dark matter. This picture explains in a natural way, why our SMCO has very
low accretion rate and why the observed angular size of the corresponding
radio source is much smaller than expected. Interactions between
dark-matter Q-balls may explain how SMCOs were seeded in galaxies and
resolve well-known problems of standard (non-interacting) dark matter.
}

\begin{document}
\maketitle
\flushbottom
\section{Introduction and summary}
\label{sec:intro}
Firm observational results indicate the existence of supermassive compact
objects (SMCOs) in galactic centers, and it is often assumed that these
objects are black holes. However, there exist observations whose
explanation requires invoking very complicated models within the
black-hole paradigm. Discussed in more detail below, they
include extremely inefficient accretion on the Milky-Way SMCO, very small
size of the resolved Galactic-center radio source, presence of SMCOs at
early stages of galactic evolution etc. Here I show that giant
nontopological solitons, Q-balls, made of a scalar field not interacting
with baryons represent a viable model for SMCOs, alternative to black
holes. This model fits well all observational data related to the
innermost part of the Milky-Way nucleus and explains its low accretion
efficiency and the small angular size of the Sgr~A$^{\star}$ radio source.
This is because the SMCO Q-balls may be considerably larger than a black
hole of the same mass, while baryonic matter penetrates the Q-ball freely
and normally passes through, and only a small fraction of baryons loses
the angular momentum in collisions with each other and gets
gravitationally trapped in the central part of the SMCO. At some stage,
this mass gain may force the Q-ball to collapse into a black hole, giving
rise to a temporal burst of the galaxy's activity. While a detailed
quantitative analysis requires complicated large-scale numerical
simulations, far beyond the scope of this note, order-of-magnitude
estimates suggest that smaller Q-balls made of the same scalar field
$\phi$ may be produced in the early Universe in the amount and with cross
sections relevant for self-interacting dark matter. They may give birth to
supermassive Q-balls in galactic centers via the gravothermal collapse,
helping at the same time to alleviate some problems of the standard,
non-interacting dark matter. The rest of dark matter may be constituted
either of the $\phi$ particles, or of similar Q-balls of different size
and, hence, of different cross section.
Future observations will help to choose between this and conventional
models.

\section{Supermassive central objects}
\label{sec:SMCO}
Thanks to intense development of observational techniques, enormous amount
of information about galactic central objects have been obtained in recent
years. The best-studied central object resides in our own Galaxy, the
Milky Way. Its observations revealed the following facts suggesting
that the central object is compact and very massive (see e.g.\
Ref. \cite{GC-review2013} for a recent review).

(1). Infrared observations of stars moving around the central object
in (almost) Keplerian elliptic orbits \cite{S2star-1, S2star-2, new}
indicate \red \cite{new} \black that the central mass is
\red $M=\left(4.02 \pm 0.20 \right)\times 10^{6}M_{\odot}$\black, where
$M_{\odot}$ is the solar mass, and that this mass is located within the
pericenter distance of the S2 star, $\approx 0.58$~mpc. Expressed in terms
of the Schwarzschild radius, $R_{S}$, of a black hole with the mass $M$,
this distance is $\sim 1500 R_{S}$. The mass of matter between the
pericenter and apocenter ($\approx 23000 R_{S}$) distances of S2 does not
exceed \red $\sim 0.01M$\black.

(2). Radio observations reveal a strong point-like source,
Sgr~A$^{\star}$, whose position coincides with the focus of the stellar
orbits within the experimental precision, dominated by a
systematic error between the infrared and radio coordinate reference
frames ($\sim 200 R_{S}$). Submillimeter observations indicate
\cite{EHT-small-image-1, EHT-small-image-2} that the angular size of the
emitting region is $\sim (30-40)~\mu$as, while the expected
apparent horizon size of a black hole of mass $M$ is $\approx 52~\mu$as,
taking into account light deflection in the black-hole gravitational field.

(3). While the stars surrounding this radio source move around it at
well-detected velocities, the apparent motion of Sgr~A$^{\star}$ itself
may be fully accounted for by the rotation of the Solar System in the
Galaxy; in particular, the source does not move, within the measurement
precision, perpendicularly to the Galactic plane \cite{properMotion1}.
This implies \cite{properMotion2} that the mass $M_{\star}$ of the body
emitting in radio is $M_{\star} \gtrsim 0.1 M$, \red unless $M_{\star}\ll
M$ and the radio source is located within $\sim 50 R_{S}$ from the center
of mass. \black

(4). The steady infrared luminosity of Sgr~A$^{\star}$ is very modest.
Compared to much stronger radio emission, which presumably originates
from the gas falling to SMCO, this is often used \cite{Broderick1,
Broderick2} as an argument that the central object cannot have a surface,
because the latter would be heated by the falling gas and shine in the
infrared $\sim 250$ times brighter.

(5). The bolometric luminosity of the SMCO is very low, $\sim 3 \times
10^{-9}L_{\rm Edd}$, compared to the Eddington luminosity $L_{\rm Edd}$,
typical for powerful active galactic nuclei and set by the balance between
the accretion flow and the radiation pressure. Partly, this may be
attributed to the lack of material for accretion: recent \textit{Chandra}
observations indicate \cite{ChandraScience} that $\sim 10^{-5} M_{\odot}$
of hot gas per year is available for accretion in the SMCO
sphere of influence. This amount of matter is insufficient to establish a
large-scale accretion disk and would correspond to the Bondi
accretion rate of $\sim 10^{-4}$ times the Eddington one. However, the
same observations reveal that less than 1\% of this captured matter falls
to the SMCO, so that the inflow of matter is almost balanced by an outflow.

The extreme compactness of the SMCO, facts (1) and (3) above, as well
as the absence of a surface interacting with matter being accreted, fact
(4), suggest that the Galactic central object may be a supermassive black
hole (SMBH), though it is presently unclear how these objects
were initially formed in galactic centers (see e.g.\
reviews \cite{Dokuchaev:origin-review, form-BH-rev2012} and
below). The fact (2), that the radio source looks
smaller than horizon, shocking at first sight because one would
expect the radio emission to come from an extended accretion disk, may be
understood (see e.g.\ Ref.~\cite{EHT-jet} and references therein) in a
model where the source is a small region at the base of a jet. However,
the jet model has serious tension \cite{JetIsBad} with polarization
measurements \cite{AccrRateLimit}.
More observational data are required to understand definitely the
origin of the radio emission.

The fact (5), that is low efficiency of accretion and radiation, may find
its explanation in a variety of complicated models of
radiatively inefficient accretion flows, see Ref.~\cite{YuanNarayanReview}
for a review. Some of them accomodate inflows and outflows similar to
those suggested by the Chandra results; determination of the actual
mechanism of accretion awaits further observations.

All these observational results, while being consistent with the SMBH
hypothesis,
do not exclude a less compact object
without a surface heated by falling baryonic matter. Alternatives to SMBHs
have been considered and it has been acknowledged that a compact object
made of a scalar field, often called a boson star \cite{RufBon}, is
perhaps
the only acceptable known candidate, see e.g.\ Ref.~\cite{Torres}.
Gravitating Q-balls discussed here represent a subclass of boson stars,
though their properties differ drastically from those of classical
boson-star solutions kept stable by gravity. The alternatives to SMBHs are
often disregarded because of the lack of observational reasons favouring
them against ``familiar'' black holes, as well as the lack of
answers to questions of how they could be formed and why they do not
collapse to black holes. All these reasons are not applicable, however, to
the Q-ball model discussed below.

\section{Q-balls as SMCOs}
\label{sec:QB-SMCO}
Nontopological solitons, or Q-balls, are compact configurations in scalar
field theories whose stability is due to the conservation of a global
charge \cite{First-Q-ball, FriedbergLeeSirlin, Coleman} (see e.g.\
a review \cite{LeePang:Phys.Rep.} and
textbooks \cite{KlassPolya, GorbyRu}). They received much attention from
researchers in field theory and cosmology, but they also have been
studied in condensed-matter
systems \cite{Q-balls-observed!}. For scalar potentials satisfying certain
conditions, the minimal-energy configuration at a fixed charge $Q$ is
compact, that is a Q-ball of charge
$Q$ always has lower energy than $Q$ free particles. This is
guaranteed by typical power-law dependence of the Q-ball mass
$M$ on its charge, $M \propto v Q^{A}$, where $v$ is some
model-dependent dimensionful parameter of the potential and $A \le 1$
(I
use the same notation, $M$, for the mass of the Galactic SMCO and for a
generic Q-ball mass; universal units, $\hbar=c=1$, are used).
The
radius of the soliton $R$ is also determined by its charge, $R \propto
v^{-1} Q^{B}$.

These nontopological solitons may be macroscopic classical
objects (see, however, an example of particle-like Q-balls in
Ref.~\cite{Emin}) and may play a role of SMCOs. If a giant Q-ball
resides in the Milky-Way center, it has to satisfy several constraints in
order to explain observational data. First, it should not be too compact
(and collapse into a black hole), but at the same time should not be too
large (and disturb the S2-star results, fact (1)). Only in a few cases
explicit solutions for gravitating Q-balls were constructed.
However, general arguments suggest \cite{Lee:critical-mass,
FriedbergLeePang:BH} (see also Ref.~\cite{SaveKillBalls}) that the Q-ball
collapses to a black hole when its radius $R$ is of order the
gravitational radius, $R_{S}=2GM$, for its mass $M$ (here $G=1/M_{\rm
Pl}^{2}$ is the gravitational constant). We therefore obtain a very
general constraint on the parameters of a putative Q-ball in the Milky-Way
center, $ R_S \le R \le 1500 R_S, $ where the last inequality involves the
pericenter distance of the S2 star. Since the $R(Q)$ and $M(Q)$
dependencies relate $M$ and $R$, this condition may be formulated in terms
of the known SMCO mass $M$ and the potential parameter $v$.

\begin{figure}
\centerline{\includegraphics[width=0.67\columnwidth]{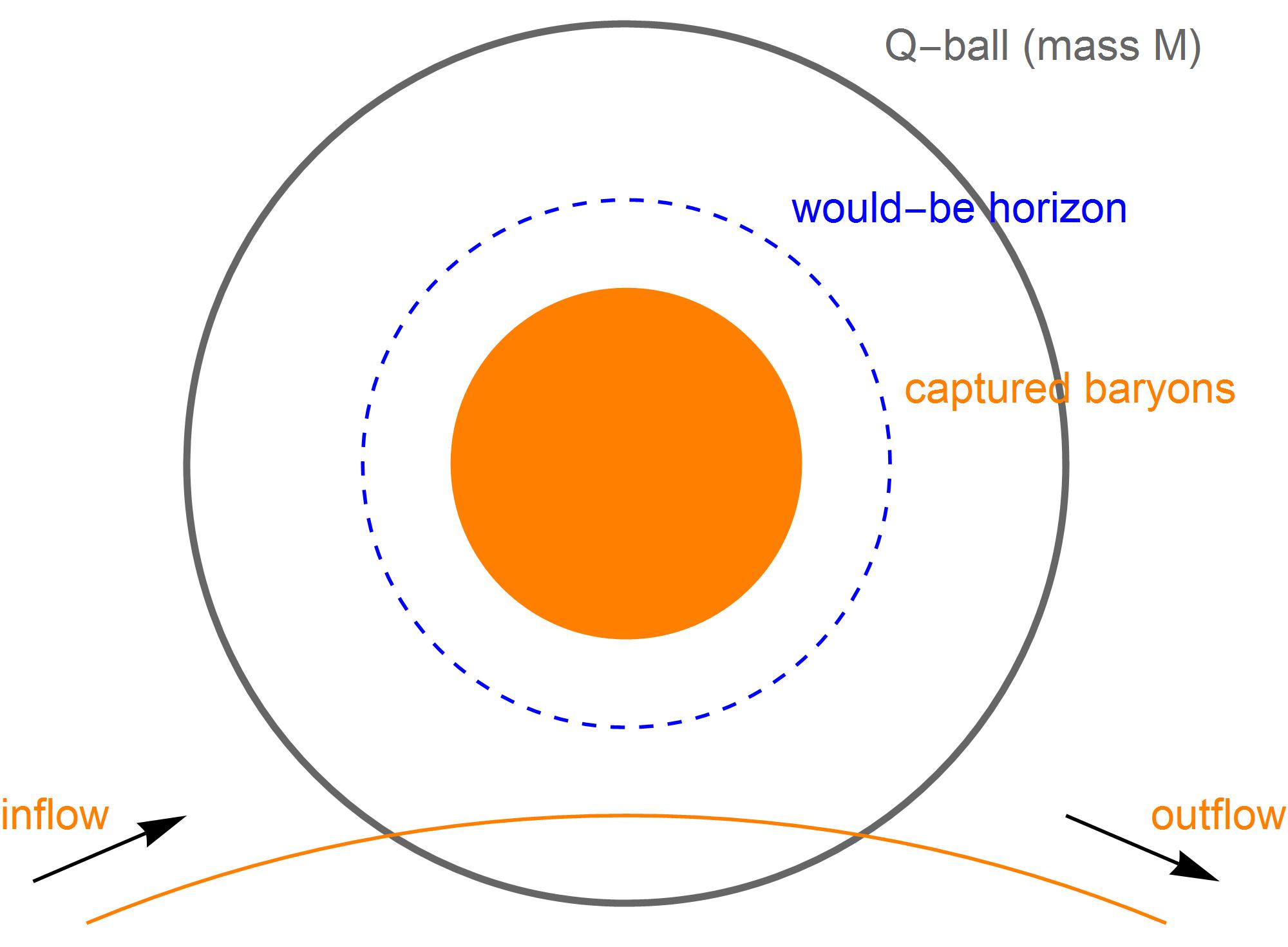}}
\caption{\label{fig:sketch}
A sketch of the proposed SMCO. The dashed line represents the horizon
which a black hole of mass $M$ would have. The actual SMCO of mass $M$,
the Q-ball, is larger. Baryonic matter may pass through it, giving rise to
inflows and outflows, while a small fraction of baryons lose their angular
momentum and are captured inside, forming a relatively small radio-emitting
blob (shaded area).
}
\end{figure}

Next, the fact (4) requires that the Q-ball does not have a surface
interacting with baryons and emitting heat. This does not mean it should
have a horizon: the field $\phi$ can simply not interact with baryons,
or have a very weak interaction. This would make it possible for the
accreting matter to penetrate inside (and to go through) the SMCO, thus
explaining the absence of thermal emission from the surface.
Suppose that the SMCO is much larger than a black hole of the same mass
(large compared to the expected size of the accretion disk, say, $R \gtrsim
100 R_{S}$). Most part of the falling
baryonic matter simply passes through the Q-ball, providing for an outflow
which balances the inflow, fact (5). A small fraction of baryons, however,
may experience collisions between themselves and lose angular momentum. It
is this matter which concentrates in the inner part of the Q-ball and
forms the radio source, whose size may be smaller than the horizon size of
the would-be black hole with the full SMCO mass, fact (2) (see
Fig.\ref{fig:sketch}).
We note that the possibility to have a
radio source of Sgr~A$^\star$ inside a Q-ball was recognized, without a
discussion, in Ref.~\cite{geodesic}, while orbits of test particles
penetrating a Q-ball were studied in Ref.~\cite{inspiral}.
\red The mass of this central baryonic blob may grow and become comparable
to the Q-ball mass, fact (3); however, such a dense configuration of
baryons might become unstable and collapse, so the second option in fact
(3) looks more probable. \black
Eventually, the mass of the
entire central object (the Q-ball plus captured baryons) exceeds the
stability limit, and the SMCO collapses into SMBH\footnote{Or into a
``Q-hairy black hole'', see Ref.~\cite{1509.02923} for solutions
and Ref.~\cite{1509.00021} for possible observational
signatures.}. One may speculate that the drastic change in the accretion
process, associated with this transition, enhances the activity of the
galaxy, which returns to a more quiscent state after stationary accretion
to a newborn black hole settles down.
A detailed study of this
prospect is beyond the scope of this note.\footnote{\red This does not
exclude other forms of less violent activity both at the earlier (Q-ball)
and later (SMBH) stages, for instance, those resulted in
Galactic-Center outflows which might be responsible for Fermi bubbles.}

We see that the Q-balls not only fit all observational constraints on the
Galactic SMCO, but may explain easily some of peculiar phenomena seen in
the Milky-Way center. I turn now to the question of how these objects might
be created in galactic centers. This would relate them to the dark
matter.

\section{Birth of supermassive Q-balls}
\label{sec:birth}
Recent observations suggest that SMCOs were present in galaxies quite
early, with $\sim 10^{9} M_{\odot}$ objects observed at $z\approx
7.1$ \cite{BH_z=7.1}, $\sim 10^{10} M_{\odot}$ at $z\approx
6.3$ \cite{BH_z=6.30} etc. The standard picture of black-hole formation
and growth, implying $\sim 10^{2} M_{\odot}$ seed black holes from stellar
explosions, accreting matter and possibly merging, faces difficulties in
explaining these observations. It has been suggested that, if dark-matter
particles are self-interacting, it helps to solve this problem,
either by enhanced accretion rate on the seed black
hole \cite{Ostriker:SIDM-BH-bu-not-seed} or by formation of the
supermassive seed itself through the gravothermal
collapse \cite{Shapiro:SIDM-BH-seed}. The first mechanism, though
efficient, becomes suppressed shortly after it starts to operate, when
dark-matter particles start to interact before falling to SMCO
\cite{2-phase-accretion}. A recent quantitative study \cite{uSIDM-BH},
based on $N$-body simulations, suggests that the second mechanism may be
viable if a small fraction ($f \lesssim 0.1$ in terms of the density
$\rho$) is ``ultra-strongly'' interacting dark matter (uSIDM). At a time
scale $\sim10^{6}$~yr, a central part of the dark-matter halo of mass
$M_{\rm halo}$ collapses to a dense object of mass $\sim 0.025f M_{\rm
halo}$, and the mechanism explains observational data (and the correlation
between the SMCO mass and the total halo mass, e.g.\
Ref.~\cite{MBH-Mhalo}) for $f \lesssim 0.1$ and uSIDM with cross section
per unit mass $\bar\sigma\equiv \sigma/M \gtrsim (0.3/f)$~cm$^{2}$/g. The
remaining $(1-f)$ of dark matter must have weaker interactions. The
value of $\bar\sigma$ was constrained, for $f=1$, from observations of the
Bullet Cluster, $\bar\sigma \lesssim 0.7$~cm$^{2}$/g \cite{Bullett}, and
from the en{\red s}emble of interacting clusters, $\bar\sigma \lesssim
0.47$~cm$^{2}$/g \cite{1503.07675}. \red However, recent detailed
simulations of the Bullet Cluster \cite{1605.04307} allow for
$\bar\sigma\sim 2$~cm$^{2}$/g, \black while data on the A320 merging system
suggest $\bar\sigma\simeq (0.94 \pm 0.06 )$~cm$^{2}$/g \cite{NewBullett}.
There are no constraints on $\bar\sigma$ for $f \lesssim 0.1$.

Q-balls may work as interacting dark matter
 \cite{KusenkoShaposhnikov, Kusenko:SIDM, Enqvist:SIDM}, but application
of results of Ref.~\cite{uSIDM-BH} to them is not straightforward. The
interaction cross sections are different for Q-balls of different charges:
up to a model-dependent factor of order one, they are  geometrical
\cite{CrossSection1, CrossSection2}, $\sigma\sim\pi R^{2}$. Since $R$
depends on $Q$, the population of Q-balls of various charges is not
exactly the system studied in Ref.~\cite{uSIDM-BH}. Next, unlike supposed
in Ref.~\cite{uSIDM-BH}, the Q-ball cross section is not purely elastic.
Refs.~\cite{CrossSection1, CrossSection2} suggest that the elastic and
inelastic cross sections are roughly equal, and processes of merging and
charge exchange are possible at approximately the same rate as scattering:
it is these processes which are responsible, in the end, for the SMCO
formation. One might expect that, from an astrophysical perspective,
account of these processes, relevant only in the dense central core of a
dark-matter halo, would make it even easier to explain the SMCO formation
and to solve the cusp-core problem \cite{Moore} simultaneously; however,
only dedicated numerical simulations may give a quantitative description
of the corresponding processes. In what follows, we apply the results of
Ref.~\cite{uSIDM-BH} to the Q-ball system, keeping in mind that this would
give order-of-magnitude estimates only. The key difference
is that the gravothermal collapse (whose start is
demonstrated but subsequent development not studied in
Ref.~\cite{uSIDM-BH}) ends now by a formation of a giant Q-ball, not a
black hole.
\red
The collapse is expected to be stopped by the inelasticity described above
which results in Q-ball merging and consequent reduction of the number of
particles in the system (in a similar way, the gravothermal collapse of a
globular cluster is stopped by formation of double, triple etc.\ stellar
systems). A detailed quantitative description of the end of the collapse
would require extensive numerical simulations. \black

A variety of mechanisms for dark-matter Q-ball production have been
proposed and studied \cite{KusenkoShaposhnikov, KolbWrong, KolbCorrected,
RuLevin, KasuyaKawasaki-gauge-med, Enqvist:simulations,
KawasakiGravMed-Qdistr}. In all of them, initial charge asymmetry either
in the scalar-field condensate or in the ensemble of scalar particles is
required, so that there exists a net charge density, subsequently
collected and trapped in Q-balls. Therefore, each newborn Q-ball collects
its charge from some volume $V$. The charge asymmetry is defined as
$\eta_{Q}=n_{\rm Q}/s$, where $n_{\rm Q}$ is the initial charge density and
$s$ is the entropy density at the moment of Q-ball formation. Suppose that
the part $\xi$ of the charge is collected into Q-balls of some typical
charge $Q$.
Then the number density of Q-balls with the typical
charge is $n\sim \xi \eta_{Q} s/Q$. This may be related to the
present-day mass density of Q-balls $\rho_{0}=M(Q)n_{0} \sim M(Q) \xi
\eta_{Q} s_{0}/Q$, where $s_{0}\sim 3 \times 10^{3}$~cm$^{-3}$ is the
present-day entropy density. On the other hand, the cross section per unit
mass is $\bar\sigma=\sigma/M(Q) \simeq \pi R(Q)^{2}/M(Q)$. Eliminating
$Q$, one obtains a relation between $\rho_{0}$ and $\bar\sigma$, the two
key dark-matter parameters.

\section{A particular model}
\label{sec:particular}
Consider explicitly a model where Q-balls are produced in the first-order
phase transition \cite{GorbyRu, RuLevin} (see    \red
Appendices~\ref{sec:general}, \ref{sec:comments-on-models} \black for
discussion of other models). It has two scalar fields, a complex one
$\phi$ which Q-balls are made of and a real one $\chi$ whose vacuum
expectation value gives mass to $\phi$. A proper choice of the potential
results in a first-order phase transition from the false-vacuum value of
$\chi$ (massless $\phi$) to the true vacuum (heavy $\phi$). The $\phi$
particles are trapped in contracting bubbles of the false vacuum and
produce Q-balls.

In this model, the mass scale $v$ is set by the potential difference
between the two vacua, $U=v^{4}$, while the $\phi$ mass in the true vacuum
is $m_{\phi}=\kappa v$ (in notations of Ref. \cite{RuLevin}, $U=\lambda
v^{4}$ and $m_{\phi}=h v$, so $\kappa = h/\lambda^{1/4}$). One has
$M=c_{M}vQ^{3/4}$ and $R=c_{R}v^{-1}Q^{1/4}$, where $c_{M}=4\pi\sqrt{2}/3$
and $c_{R}=1/\sqrt{2}$.
The Sgr~A$^{\star}$ constraint discussed above \red ($R_{S} \lesssim R
\lesssim 1500 R_{S}$) \black requires 1.3~keV$\lesssim v \lesssim $180~keV,
with values $v \lesssim 6$~keV favored for explanation of the weak
accretion \red ($R \gtrsim 100 R_{S}$). \black

For the Q-balls to play the role of uSIDM, there are two conditions: $f
\equiv \rho_{0}/\rho_{\rm DM} \lesssim 0.1$, where $\rho_{\rm DM} \sim
10^{-6}$~GeV/cm$^{3}$ is the present-day dark-matter density, and
$\bar\sigma f \gtrsim 0.3$~cm$^{2}$/g. The two constraints may be
satisfied simultaneously for a certain choice of parameters $(\xi,
\eta_{\rm Q})$ provided $v (\xi \eta_{\rm Q})^{1/4} \lesssim 100$~keV,
which agrees well with the Sgr~A$^{\star}$ constraints.

This model allows also for two interesting possibilities to include the
remaining $(1-f)$ fraction of the dark matter. First, the remaining
$(1-\xi)$ excess of charge, not trapped into Q-balls, is kept in the form
of $\phi$ particles. They are stable because they are the lightest
particles charged under the global $U(1)$ symmetry responsible for the
Q-ball stability. Depending on their mass and interactions, they may
represent the dominant part of the present dark-matter density.

Second,
Q-balls of different sizes are produced, and, since the cross
section of a Q-ball depends on its charge, their cross sections also vary.
The bulk of produced Q-balls may form the standard dark matter, while a
small fraction of them (more precisely, those Q-balls which contribute a
small fraction to $\rho$) play the role of uSIDM.
It is possible to estimate, \red see Appendix~\ref{sec:distr-in-Q}, \black
the distribution of the Q-balls in $Q$; while small Q-balls are born more
frequently, they carry a minor fraction of $\rho$. Requiring $\rho_{0}\sim
\rho_{\rm DM}$ and $\bar\sigma \lesssim 1$~cm$^{2}$/g, one obtains the
bound $v \gtrsim 240$~keV. Recalling that the cosmological constraint was
a rough order-of-magnitude estimate only, we conclude that the model with
$v\sim 100$~keV may be capable of producing the dark-matter Q-balls and
the Q-ball Milky-Way SMCO at the same time. However,
\red these values of $v$ correspond to the Q-ball
of radius $R \lesssim 10 R_{S}$, and it remains to be studied how
the accretion process on such a dense object would differ from that on
 black hole.        \black

Clearly, detailed quantitative studies, which might require
full-scale computer simulations, are necessary to fully understand details
of birth and subsequent evolution of the dark-matter Q-ball system.
However, our order-of-magnitude estimates demonstrate that this
interesting scenario may be viable.

\section{Future tests}
\label{sec:future}
It would be difficult to test the proposed scenario by direct searches for
the dark-matter Q-balls, e.g. \cite{IceCubeBaksanKawasaki}: their number
density in the Universe is very low while the interaction with normal
matter is very weak, if any. Detailed studies in particular models might
reveal observable signatures of intermediate-mass Q-balls, if they are
produced. Model-dependent signatures may be found also for
gravitational-wave astronomy, coming both from the primordial formation of
Q-balls \cite{grav-waveKu, 1501.01217} and from the SMCO \cite{inspiral}.
However, definitive tests will be provided by high-resolution observations
of the Milky-Way central object.

For instance, when the G2 dusty object \cite{G2} has been discovered
on its way to the Milky-Way SMCO, orders-of-magnitude increase in the
accretion rate, and a corresponding burst of luminosity of Sgr~A$^{\star}$,
were predicted. This extended object has been already observed after
passing the pericenter, tidally disturbed but without any sign of
increased accretion \cite{1407.4354, 1411.0746, 1502.06534, 1503.08982}.
Further observations will help to understand its nature and to
shed light on the details of the accretion process.

\red
If the mass  of Sgr~A$^{\star}$ radio source is small compared to the SMCO
mass, it should move within the SMCO, and multi-epoch astrometric
observations may discover this motion, which has not been seen yet because
of insufficient precision.
\black

The Event
Horizon Telescope \cite{EHT}, an Earth-size radio interferometer working
at
the \red wavelength \black of 1.3~mm, might be able to resolve the
black-hole shadow, if the Milky-Way SMCO is a black hole, in a few years.
Future ``Millimetron'' spaceborn interferometer \cite{Millimetron} would
be able to have horizon-scale resolution for dozens of nearby SMCOs, and
its observations would establish the nature of supermassive objects in
galactic centers.

\appendix
\section{Distribution of Q-balls in $Q$ after the first-order
phase transition}
\label{sec:distr-in-Q}
Let us estimate the distribution of
Q-balls in $Q$ to see that, while small Q-balls are born more frequently,
they carry a minor fraction of $\rho$.
The charge of a newborn Q-ball
is determined by the volume of the remaining bubble of the old phase. The
size of suitable bubbles may be estimated from the condition that a
smaller bubble of the new phase is not created inside before the bubble
collapses. The probability that a new-phase bubble is created in the
volume of the old-phase bubble $V$ is proportional to $V$ times the time
of the bubble collapse, that is to $V\times V^{1/3}=V^{4/3}\propto
Q^{4/3}$. Therefore, the probability to create a Q-ball with charge $Q$,
and hence the number density of these Q-balls (the number of Q-balls with
the charge $Q$ per unit volume),
$$
n(Q) \propto 1-\left(\frac{Q}{Q_\star}   \right)^{4/3},
$$
where $Q_{\star}$ is some characteristic (maximal) charge estimated in
Ref.~\cite{RuLevin}. This expression works for $Q \gtrsim Q_{\rm min}$,
where $Q_{\rm min} \ll Q_{\star}$ is the minimal charge of a stable
Q-ball. This means that most of the Q-balls are born with small charges,
$Q\sim Q_{\rm min}$. However, their contribution to the mass density is
small,
$$
\rho(Q) \propto M(Q)n(Q) \propto Q^{3/4} \left(1- \left(Q/Q_\star
\right)^{4/3} \right).
$$
This function has a maximum at $Q\sim 0.5 Q_{\star} \gg Q_{\rm min}$, and
most of the density is carried by these large Q-balls with relatively
small $\bar\sigma$.

Let us divide artificially the Q-balls born after the phase transition
into two populations; those with $Q_{\rm min} \le Q \le Q_{0}$, for some
$Q_{0}$, would represent uSIDM, while those with $Q_{0} \le Q \le
Q_{\star}$ would be the bulk dark matter. Then, we should simultaneously
satisfy the following constraints:

-- the value of $\bar\sigma f$, summed over uSIDM, $\langle \bar\sigma f
\rangle_{\rm u} \gtrsim 0.3$~cm$^{2}$/g;

-- the fraction of uSIDM $\langle f \rangle_{\rm u} \lesssim 0.1$;

-- the mean cross section of the bulk dark matter $\langle \bar \sigma
\rangle_{\rm b} \lesssim 1$~cm$^{2}$/g.

Replacing sums over $Q$ by integrals, one finds
$$
\langle \bar\sigma f \rangle_{\rm u}
=
\bar\sigma_\star
\frac{ \int_{x_{\rm min}}^{x_0} x^{-1/4} x^{3/4} (1-x^{4/3})\, dx}%
{\int_{x_{\rm min}}^{1}            x^{3/4} (1-x^{4/3})\, dx},
$$
where
$x=Q/Q_{\star}$,
$x_{0}=Q_{0}/Q_{\star}$,
$x_{\rm min}=Q_{\rm min}/Q_{\star}$
and
$\bar\sigma_{\star} = \bar\sigma(Q_{\star})$.
For
$x_{\rm min} \ll x_{0} \ll 1$,
one has
$ \langle \bar\sigma f \rangle_{\rm u} \approx 2.7
x_{0}^{3/2} \bar\sigma_{\star} $.
Similarly,
$$
\langle f \rangle_{\rm u}=
\frac{\int_{x_{\rm min}}^{x_0} x^{3/4} (1-x^{4/3})\, dx}%
{\int_{x_{\rm min}}^{1}  x^{3/4} (1-x^{4/3})\, dx}
\approx 2.3 x_0^{7/4}
$$
and
$$
\langle \bar\sigma \rangle_{\rm b}
=
\bar\sigma_\star
\frac{
\int_{x_{0}}^{1} x^{-1/4} x^{3/4} (1-x^{4/3})\, dx
}{
\int_{x_{\rm min}}^{1}  x^{3/4} (1-x^{4/3})\, dx
}
\approx (1.3-2.7 x_0^{3/2})
\bar\sigma_\star
.
$$
For \red $x_{0}\sim 0.2$ and
$
\bar\sigma_\star
\sim 1$~cm$^{2}$/g, \black all three conditions are \red marginally \black
satisfied. As is shown in the main text, for $v\sim 100$~keV, correct
values of $\bar\sigma$, which we required here, correspond to the correct
density of dark matter while Sgr~A$^{\star}$ constraints are also
satisfied. \red All the relations used are approximate only, so their
approximate solution is satisfactory. However, this means that the model
should be quite fine-tuned for this scenario. The case when the bulk of
dark matter is made not of Q-balls but of $\phi$ particles remains quite
general. \black

\section{General models}
\label{sec:general}
Numerous potentials allowing for Q-balls have been
suggested. To the best of author's knowledge, all of the studied
models lead to particular values of the power-law exponents $A, B$ in
$M\propto Q^{A}$, $R\propto Q^{B}$ relations, namely, $A=(6-\alpha
)/(2(4-\alpha ))$ and $B= (2-\alpha )/(2(4-\alpha ))$ for a certain
$\alpha $, $-2\le\alpha\le 2$.

The models with $0\le \alpha\le 2$ are relevant for the case of ``almost
flat'' potentials which behave as $|\phi|^{\alpha}$ at large $|\phi|$,
see e.g.~\cite{KusenkoShaposhnikov}. It is easy to obtain explicit
expressions for the general case.

Mass:
$$
M(Q)=
c_M v
Q^\frac{6-\alpha }{2(4-\alpha )},
$$
where
$$
c_M=2\pi
\frac{4-\alpha }{3-\alpha }
\left(\frac{(3-\alpha)}{12} \right)^{\frac{1}{4-\alpha }}.
$$

Radius:
$$
R(Q)=
c_R v^{-1}
Q^\frac{2-\alpha }{2(4-\alpha )},
$$
where
$$
c_R= \frac{1}{2}
\left(\frac{(3-\alpha)}{12} \right)^{-\frac{1}{4-\alpha }}.
$$

The Sgr~A$^{\star}$ constraint:
$$
\frac{F}{1500} \lesssim \frac{v}{M_{\rm Pl}} \lesssim  F,
$$
where $M_{\rm Pl}\approx 1.2 \times 10^{19}$~GeV is the Planck mass and
$$
F=
c_M^{-\frac{2-\alpha }{2(4-\alpha )}}
c_R^{\frac{6-\alpha }{2(4-\alpha )}}
\left(\frac{M}{M_{\rm Pl}} \right)^{-\frac{2}{4-\alpha }}.
$$
Cross section per unit mass:
$$
\bar\sigma=c_\sigma v^{-3} Q^{-\frac{2+\alpha }{2(4-\alpha) }},
$$
where
$$
c_\sigma=
\frac{3-\alpha }{4-\alpha }
\left(\frac{(3-\alpha)}{12} \right)^{-\frac{3}{4-\alpha }}
$$
(note that
$\bar\sigma$ always decreases with $Q$).

Density -- cross section relation:
$$
\rho_0=c_M v \left(\frac{\bar\sigma v^3}{c_\sigma}
\right)^{\frac{2-\alpha }{2+\alpha }} \eta_{Q} s_0.
$$

Note that the $\alpha=0$ case also corresponds to a completely different
model of Refs.~\cite{FriedbergLeeSirlin, RuLevin} discussed in the main
text.
The values of $c_{M}$, $c_{R}$ and $c_{\sigma}$ for the model of
Ref.~\cite{FriedbergLeePang:BH} with $\alpha =-2$  are slightly different,
$c_{M}=3\pi$, $c_{R}=1/2$, $c_{\sigma}=1/12$, but the difference in
numerical constraints cannot be seen by eye in
further plots.

Consider two scenarios.

(i). The bulk of produced Q-balls forms the standard dark matter, while a
small fraction of them (more precisely, those Q-balls which contribute a
small fraction to $\rho$) play the role of uSIDM. Then, one should require
$\rho_{0}\sim \rho_{\rm DM} \sim 10^{-6}$~GeV/cm$^{3}$ and $\bar\sigma
\lesssim 1$~cm$^{2}$/g. Since $\eta_{Q}<1$ by definition, we obtain a
{\em lower} limit on $v$.

(ii). The bulk of Q-balls forms uSIDM while most of the dark matter is
represented by a completely different component. This requires $\rho_{0}=f
\rho_{\rm DM}$ with $f \lesssim 0.1$ and $\bar\sigma f \gtrsim
0.3$~cm$^{2}$/g. The resulting {\em upper} limits on $v$  depend now on the
assumed $\eta_{Q}$.

\begin{figure}
\centerline{\includegraphics[width=0.67\columnwidth]{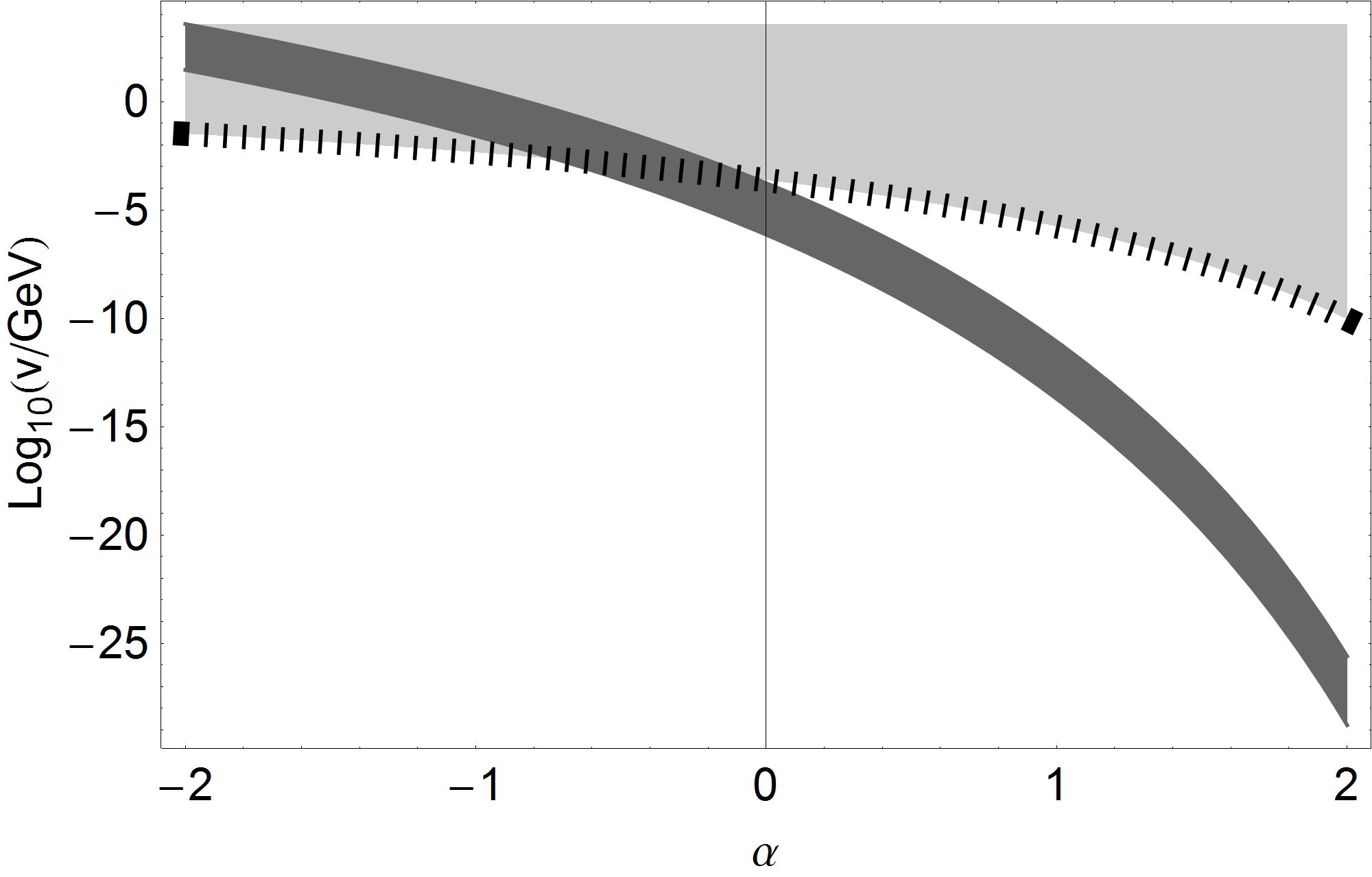}}
\caption{\label{fig:llim}
Constraints on the scale parameter $v$ of the scalar potential for the
case when all dark matter is made of Q-balls. The dark gray band
represents the values required for a Q-ball SMCO in the Milky Way. The
light-gray shaded area is allowed by constraints derived from the
dark-matter density and cross section. The latter constraints are order-of
magnitude estimates. A $f \lesssim 0.1$ fraction of Q-balls interacting
considerably stronger and responsible for SMCO formation is allowed. }
\end{figure}

\begin{figure}
\centerline{\includegraphics[width=0.67\columnwidth]{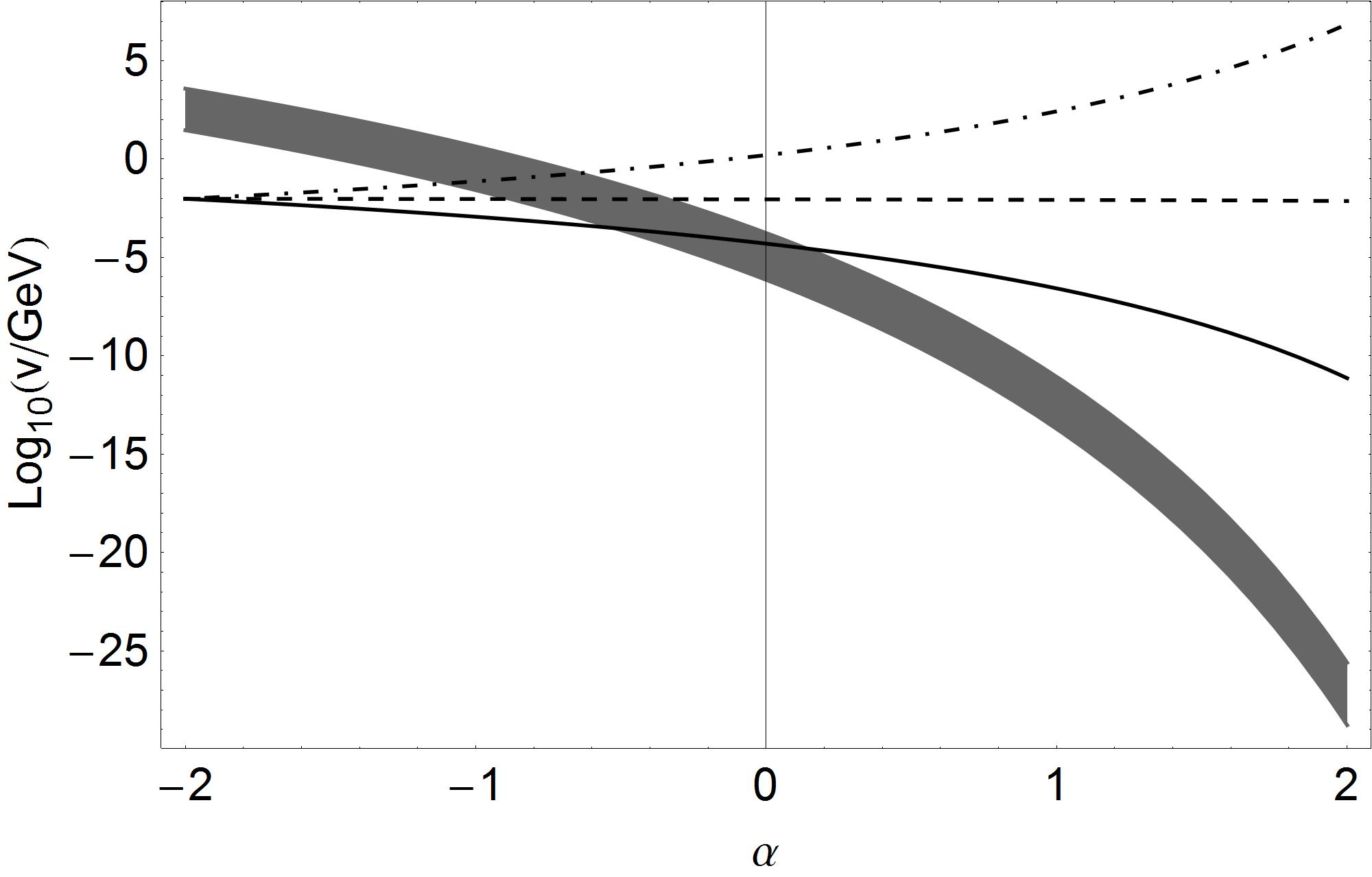}}
\caption{\label{fig:ulim}
Constraints on the scale parameter $v$ of the scalar potential for the
case when Q-balls represent only the fraction of dark matter responsible
for SMCO formation by gravothermal collapse. The dark gray band represents
the values required for a Q-ball SMCO in the Milky Way. Thin lines
represent \textit{upper} limits on $v$ for the charge asymmetry
$\eta_{Q}=1$ (solid), $10^{-9}$ (dashed) and $10^{-18}$ (dash-dotted).
The latter constraints are order-of magnitude estimates.
}
\end{figure}

Figures \ref{fig:llim}
and \ref{fig:ulim}
illustrate constraints on
scenarios (i) and (ii), respectively. The dark
grey band in both plots represents the values required by the
Sgr~A$^{\star}$ constraints. In Fig.~\ref{fig:llim}, the area of
parameters allowed for scenario (i) is shown as a light grey region.
Constraints on scenario (ii) cannot be
formulated unless a particular value of the charge asymmetry $\eta_{Q}$ is
assumed; they put upper limits on $v$ shown in Fig.~\ref{fig:ulim} for
various values of $\eta_{Q}$.

\section{Comments on particular models}
\label{sec:comments-on-models}
(A). Flat scalar potential, $U(\phi)\sim {\rm const}$ at large $|\phi|$,
$\alpha=0$ \cite{Coleman, KusenkoShaposhnikov}. Q-balls form from the
$\phi$ condensate which develops instabilities.
This process is highly nonlinear,
and numerical simulations are required to study it. It has been shown
\cite{KasuyaKawasaki-gauge-med} that most part of the charge is collected
to large Q-balls (low $\bar\sigma$), while a number of small Q-balls
(large $\bar\sigma$) are also produced. This mechanism may, in principle,
work in scenario (i), though quantitative results on the distribution of
produced Q-balls in $Q$ are presently missing.

(B). Second-order phase transition, $\alpha =0$~\cite{KolbWrong,
KolbCorrected}.
The probability to create a Q-ball is determined by fluctuations and is
exponentially small. As a result, only a small fraction of charge is
collected to Q-balls with $Q\sim Q_{\rm min}$, with exponentially
suppressed chance to create a larger Q-ball. This model may work only in
scenario (ii) because almost all Q-balls have the same size and hence the
same cross section.

(C). \red Quadratic potential with logarithmic corrections, \black $\alpha
=2$ \cite{Enqvist:simulations, KawasakiGravMed-Qdistr}. The production of
these Q-balls was studied numerically in more detail and the distribution
of Q-balls in $Q$ was presented in Ref.~\cite{KawasakiGravMed-Qdistr}.
However, in this model, $R=$const and does not depend on $Q$, while $M
\propto Q$. As a consequence, only a very light scalar field may prevent
collapsing of a massive Q-ball to a black hole, cf.\ Figs.~\ref{fig:llim},
\ref{fig:ulim} (scenario (ii) is allowed for $m\sim 10^{-26}$~GeV and
scenario (i) is excluded).

(D). Model of Ref.~\cite{FriedbergLeePang:BH}, $\alpha=-2$. Here,
$M\propto Q^{2/3}$ and $R \propto Q^{1/3}$, so $\bar\sigma$ does not
depend on $Q$. This might work for scenario (ii) only, \red because all
Q-balls have the same $\bar\sigma$ and there is no place for two
populations. However, this model \black is excluded by Fig.~\ref{fig:ulim}.
Note also that no mechanism to produce Q-balls in the early Universe is
known for this model, best studied in the context of boson stars.

\acknowledgments
I am indebted to D.~Gorbunov, A.~Kanapin, M.~Libanov, M.~Pshirkov,
G.~Rubtsov, S.~Sibiryakov and especially to E.~Nugaev and V.~Rubakov for
interesting and helpful discussions, and to Yu.~Filippov for bringing the
G2 story to my attention. I thank CERN (TH division) for hospitality at
the final stages of this work. This work is supported by the Russian
Science Foundation, grant 14-22-00161.




\end{document}